\documentclass[11pt, conference, final]{ieeetran}
\usepackage{times,epsfig,latexsym}
\usepackage{latexsym,url,epsfig}
\usepackage{enumerate}
\usepackage{umlaut}
\usepackage{theorem}
\usepackage{floatflt}
\usepackage{graphicx}
\usepackage{color}
\usepackage{calc}
\usepackage{amsmath}
\usepackage{clrscode}

\newtheorem{theorem}{Theorem}[section]

\title{Defragmenting the Module Layout of a Partially Reconfigurable
  Device
}
\author{ 
  \authorblockN{ 
    Jan C. van der Veen\authorrefmark{1}, S\'andor P. Fekete\authorrefmark{1}, 
    Mateusz Majer\authorrefmark{2}, Ali Ahmadinia\authorrefmark{2}, \\
    Christophe Bobda\authorrefmark{2}, Frank Hannig\authorrefmark{2}, and J\"urgen Teich\authorrefmark{2}
  }
  \\
  \authorblockA{ \authorrefmark{1} Department of Mathematical
    Optimization,
    Braunschweig University of Technology, Germany \\
    Email: \{j.van-der-veen, s.fekete\}@tu-bs.de }
  \\
  \authorblockA{ \authorrefmark{2} Department of Computer Science 12,
    University of Erlangen-Nuremberg, Germany \\
    Email: \{majer, ahmadinia, bobda, teich\}@cs.fau.de } }

\begin{document}

\maketitle

\begin{abstract}
  Modern generations of field-programmable gate arrays (FPGAs) allow for partial
  reconfiguration. In an online context, where the sequence of modules
  to be loaded on the FPGA is unknown beforehand, repeated insertion
  and deletion of modules leads to progressive fragmentation of
  the available space, making defragmentation an important issue.
  We address this problem by propose an online and an offline component
  for the defragmentation of the available space. 

  We consider defragmenting the module layout on a reconfigurable
  device. This corresponds to solving a two-dimensional strip packing
  problem. Problems of this type are NP-hard in the strong sense, and previous
  algorithmic results are rather limited. Based on a graph-theoretic
  characterization of feasible packings, we develop a method that can
  solve two-dimensional defragmentation instances of practical size
  to optimality. Our approach is validated for a set of benchmark instances.
\end{abstract}

\bigskip
{\bf Keywords:} Reconfigurable computing, partial reconfiguration,
defragmentation, two-dimensional packing, NP-hard problems, exact algorithms.

\section{Introduction}



One of the cutting-edge aspects of modern reconfigurable computing is
the possibility of {\em partial} reconfiguration of a device: Ideally,
a new module can be placed on a reconfigurable chip whithout
interfering with the processing of other running tasks. 
(See the end of this subsection for some pratical restrictions 
in current generations of FPGAs.) Clearly, this
approach has many advantages over a full reconfiguration of the whole
chip. Predominantly it lessens the bottleneck of reconfigurable
computing: reconfiguration time.

On the other hand, partial reconfiguration introduces a new
complexity: management of the free space on the FPGA. In the 2D model
this is an NP-hard optimization problem. There has been a considerable
amount of work to solve this problem computationally. However, due to
its computational complexity most recent work has focused on the
online setting or on the 1D area model (see \cite{swp-osrep-04} for a
recent survey).

\begin{figure}[htb]
  \centering
  \includegraphics[width=.6\columnwidth]{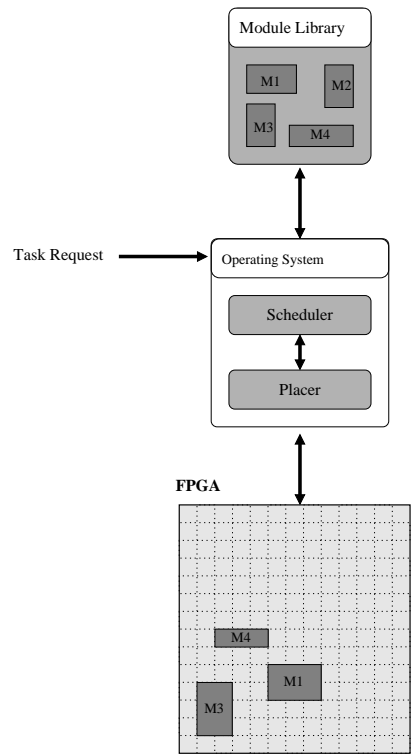}
  \caption{A schematic overview of an operating system for reconfigurable computers. Relocatable, presynthesized modules that are constrained to a rectangular layout are stored in a module library. As requests for tasks arrive, a module capable of running the task is selected, scheduled and eventually placed on the FPGA.}
  \label{fig:opsys}
\end{figure}

Management of free space and scheduling of arriving tasks are the core
components of an operating system for reconfigurable platforms (see
Figure~\ref{fig:opsys}). In all previous work these components use
simple online strategies for the placement problem. The use of these
strategies leads to fragmentation of the free space, as modules are
placed on and removed from the chip area.  This leads to situations
where a new module has to be rejected by the placer because there is
no free rectangle that could accomodate the new module even though the
total free space available would be more than sufficient (see
\cite{demss_dstpr-00} for a discussion).

In this paper we propose a different placer module. Instead of just
relying on online strategies our placer has an additional offline
component: the defragmenter. Consider the following scenario: A car is
equipped with a multimedia device that contains a partially
reconfigurable FPGA. This multimedia device is responsible for audio,
video, telephony and WLAN. While the car is in use, the device is busy
and tasks must be scheduled and modules must be placed as they arrive.
However, the recurring idle times of the car (i.e., over night) can be
utilized to optimally defragment the FPGA chip area.

This optimal defragmentation follows two goals. One is to maximize the
available contiguous free space. The other comes from the FPGA device
we use. The current XILINX Virtex-II series does not admit full
two-dimensional partial reconfiguration \cite{x-vpfcd-04}. Instead,
configuration can only be performed columnwise: While a column is
reconfigured, all other modules that use this column have to be
stopped, because reconfiguration interferes with the running tasks in
a non-trivial way. So the other goal of the offline defragmenter is to
free as many columns as possible. This way the next modules placed by
an online placer will not interfere with other modules.

The rest of the paper is organized as follows. In the next section we
describe our FPGA model and conclude that the offline optimization
problem that is to be solved is the two-dimensional strip packing
problem. In sections \ref{sec:2DSP} and \ref{sec:opp} we describe our
algorithm for solving this problem to optimality. Then we will report
on computational results. In our conclusion we hint at possible
extensions of our model.


\section{Column-oriented cost function} 

Due to its wide-spread use, our device model closely resembles that of
a XILINX Virtex-II FPGA. In our model the FPGA consists of a certain
number of reconfigurable units called {\em configurable logic blocks}
(CLBs). These CLBs are organized in $W$ columns and $H$ rows. There is
no way to reconfigure CLBs individually: Reconfiguration takes place
on the column level. We assume that it takes $c$ units of time to
configure one column of CLBs. 

On this FPGA we execute a certain set of tasks $T = \{t_1, t_2,
\ldots\}$. In an offline setting we would assume that for each task
$i$ its arrival time $a_i$ is known in advance. Some tasks may carry a
deadline $d_i$. A deadline is the time when task $i$ is required to
have finished its execution. If a task has no deadline this is
indicated by setting $d_i = \infty$.  Inter-task dependencies are
modeled by $p : T \rightarrow 2^T$, describing the predecessors
of any task.

Tasks can be executed in hardware or in software. We assume that for
each task there is at least one hard- or software module. A hardware
module is a relocatable presynthesized digitial circuit that has been
constrained to a rectangular area. In the following $w_j$ and $h_j$
denote the width and the height of the $j$-th module. As a
consequence, placing module $j$ on the FPGA takes time $c w_j$.  A
software module is a precompiled executable that can be executed, e.g., 
on a soft-core IP such as the MicroBlaze soft-cores for the XILINX
devices. For ease of notation we assume that a software module $j$
requires the width and height of its processor IP core. The set of all
modules is given by $M = \{m_1, m_2, \ldots\}$ including possible
processor cores. If a task $i$ is executed on module $j$, its
execution time is given by $e_i^j$.  In addition, each module $j$ has
a usage count $u_j$ that will be explained later.

Currently, communication between modules is still an
issue.  But as chip size and complexity increases circuit as well as
packet-based on-chip communication networks, such as DyNoC
\cite{bmkat-dnacr-04} become more and more realistic. Here we assume
the availability of a fine-grained underlying communication
infrastructure supporting intermodule communication requests.

In an offline setting we simultaneously seek for:
\begin{itemize}
\item A feasible schedule for the tasks.  In other words, each task
  $i$ is assigned a starting time $s_i$.
\item An assignment $m : T \rightarrow M$ of tasks to modules. By
  $m(t_i)$ we denote the module task on which $i$ will be executed.
\item A configuration schedule for the modules. Each module $j$ is
  assigned a configuration time $c_j$. Of course configuration and
  starting time are related through $s_i \geq c_{m(t_i)} + c
  w_{m(t_i)}$.
\item A feasible placement of the modules on the FPGA. For each module
  $j$ its location $x_i \in [0, W - w_i)$ and $y_i \in [0, H - h_i)$
  has to be determined.
\end{itemize}
Among all feasible solutions we select one that minimizes the {\em
  makespan}, i.e., the completion time of the last task. This alone is an
NP-hard optimization problem, as it contains two-dimensional packing
as a subproblem. At the same time, this problem is closely related to scheduling problems. (See \cite{sched} for an overview of classical ``one-dimensional''
scheduling problems.)

In the two-dimensional placement model, columnwise reconfiguration has
the drawback that reconfiguring a column of the FPGA affects all
modules using this column in a non-trivial way. In our model we assume
that the reconfiguration of one column interrupts all modules using this column
for the reconfiguration time $c$. Therefore, a task running on a
module $j$ is interrupted for $c |[x_j, x_j + w_j) \cap [x, x + w_i)|$
time units, if module $i$ is placed starting at column $x$.

\begin{figure}[htbp]
  \begin{center}
    \includegraphics[width=\columnwidth]{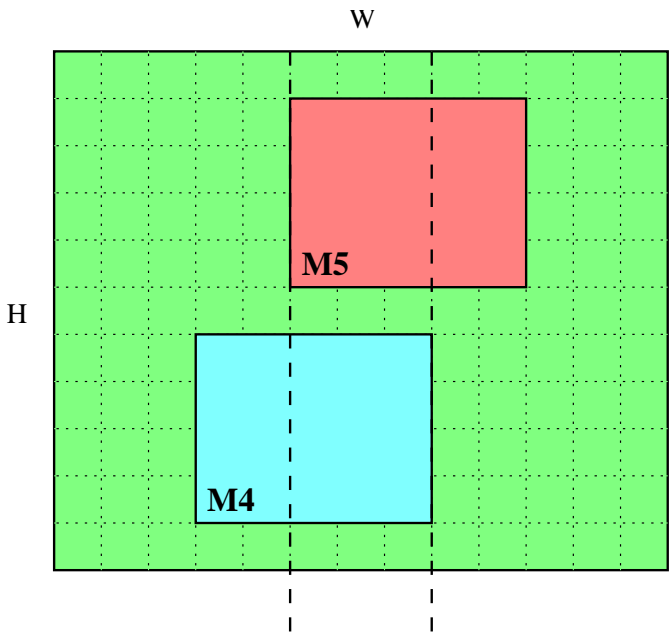}
    \caption{An FPGA of width $W = 13$ and height $H = 11$. Assume that module M4 of width $w_4 = 5$ and height $h_4 = 4$ is located at position $(3, 1)$. If module M5 of same width and height is placed at position $(5, 6)$ the resulting overlap is $3$ columns as indicated by the dashed lines. Consequently M4 is interrupted for $3c$ time units.}
    \label{fig:def}
  \end{center}
\end{figure}

There is some experimental evidence that an online placement strategy
should take this interference into account. As we showed in
\cite{at-sopxf-03}, the least interference fit (LIF) online strategy
is quite useful in this setting: New modules are placed in consecutive
columns that are used by as few other modules as possible. But in the
long term LIF faces two problems:
\begin{enumerate}
\item Free space fragmentation: Even though the free space available
  on the FPGA would allow executing a task on a hardware module
  (resulting in better quality and/or faster execution), the largest
  free space fragment available may not be able to accomodate the
  respective module.
\item Interference: Even though respecting the number of interrupted
  modules, LIF still has to interrupt modules in the long run.
\end{enumerate}

In this paper we propose a strategy that can increase the long-term
quality of the LIF strategy. As described above, our scenario gives
rise to times where the system is rather busy. On the other hand,
there also are times when the system is more or less offline or
unused. These are times when the FPGA could be defragmented. By
defragmentation we mean removing modules that have a low usage count
and then moving all modules so that a maximal number of columns is
unused. This increases the effectiveness of online strategies like
LIF.

Defragmentation as described in the paragraph above can be regarded as
the two-dimensional strip packing problem. In the next section we will
take a closer look at this classic NP-complete optimization problem.
As it turns out, for currently relevant numbers of modules, optimal
placements can still be computed, using a cutting-edge algorithm for
higher-dimensional packing.


\section{Two-Dimensional Strip Packing}\label{sec:2DSP}

Packing rectangles into a container arises in many industries,
whenever steel, glass, wood, or textile materials are to be cut, 
but it also occurs
in less obvious contexts, such as
machine scheduling or optimizing the layout of advertisements in newspapers.
The three-dimensional problem is important for practical
applications such as container
loading or scheduling with partitionable resources.
For many of these problems, objects must be positioned with
a fixed orientation; this requirement also arises when configuring
modules on a chip area.

Different types of objective functions for multi-dimensional
packing problems have been considered. 
The \emph{Strip Packing Problem}
(SPP) is to minimize the width $W$ of a strip of fixed height $H$ 
such that all rectangles fit into a rectangle of size $W\times H$.
The {\em orthogonal knapsack problem} (OKP)
requires selecting a most valuable subset $S$ from a given set of
rectangles, such that $S$ can be packed into the large rectangle.
The {\em orthogonal bin packing problem} (OBPP) considers the scenario
in which a supply of containers of a given size is given and the objective
is to minimize the number of containers that are needed for
packing a set of boxes.

Crucial for all those optimization problems is the corresponding decision
problem:
The \emph{Orthogonal Packing Problem} (OPP)
is to decide whether a given set of rectangles can be placed within a given
rectangle of size $W\times H$.  
As all of the above problems can be generalized to arbitrary dimensions,
we denote by SPP-$d$, OKP-$d$, OBPP-$d$, and OPP-$d$ the
strip-packing problem, the orthogonal knapsack problem, the orthogonal
bin packing problem, and the orthogonal packing problem, respectively,
in $d$ dimensions. (E.g., when considering scheduling problems
on an FPGA implies considering two space and one time dimension,
yielding $d=3$.)
Being a generalization of the one-dimensional problem {\sc 3-Partition},
the OKP-$d$ is NP-complete in the strict sense, and so the
corresponding optimization problems are NP-hard \cite{GJ79}.

Dealing with an NP-hard problem (often dubbed ``intractable'')
does not mean that it is impossible to find provably optimal solutions.
While the time for this task may be quite long
in the worst case, a good understanding of the underlying mathematical
structure may allow it to find an optimal solution (and prove its optimality)
in reasonable time for a large number of instances. 
A good example
of this type can be found in \cite{GRO80}, where the exact solution
of a 120-city instance of the Traveling Salesman Problem
is described. In the meantime, benchmark instances
of size up to 13509 and 15112 cities have been solved to
optimality \cite{13509},
showing that the right mathematical tools and sufficient computing power
may combine to explore search spaces of tremendous size. In this sense,
``intractable'' problems may turn out to be quite tractable.

Higher-dimensional packing problems have been considered by a great
number of authors, but only few of them have dealt with the exact
solution of general two-dimensional problems.  See \cite{ESA,pack1}
for an overview.  It should be stressed that unlike one-dimensional
packing problems, higher-dimensional packing problems allow no
straightforward formulation as integer programs: After placing one box
in a container, the remaining feasible space will in general not be
convex. Moreover, checking whether a given set of boxes fits into a
particular container is trivial in one-dimensional space, but NP-hard
in higher dimensions.

Nevertheless, attempts have been made to use standard approaches of
mathematical programming.  Beas\-ley~\cite{Bea85} and
Hadjiconstantinou and Chris\-to\-fides~\cite{HC95} have used a
discretization of the available positions to an underlying grid to get
a 0-1 program with a pseudopolynomial number of variables and
constraints. Not surprisingly, this approach becomes impractical
beyond instances of rather moderate size.  

To our knowledge there is only one work that tries to solve SPP to
optimality. In \cite{mmv-easpp-03} the authors derive improved lower
and upper bounds for the two-dimensional strip-packing problem. These
bounds are based on a continuous relaxation of the one-dimensional
contiguous bin-packing problem (1CBP). These bounds are used in a
branch-and-bound type algorithm to solve 27 benchmark instances from
the literature.

In~\cite{ESA,pack1,pack2,pack3,tfs-ohrt-01}, a different approach to
characterizing feasible packings and constructing optimal solutions is
described.  A graph-theoretic characterization of the relative
position of the boxes in a feasible packing (by so-called
\emph{packing classes}) is used, representing $d$-dimensional packings
by a $d$-tuple of interval graphs (called \emph{component graphs})
that satisfy two extra conditions. This factors out a great deal of
symmetries between different feasible packings, it allows to make use
of a number of elegant graph-theoretic tools, and it reduces the
geometric problem to a purely combinatorial one without using
brute-force methods like introducing an underlying coordinate grid.
Combined with good heuristics for dismissing infeasible sets of
boxes~\cite{IPCO}, a tree search for constructing feasible packings
was developed.  This exact algorithm has been implemented; it
outperforms previous methods by a clear margin.
This approach has been extended to strip-packing problems in the presence
of order constraints; see \cite{fkt-mdpoc-01}.
(Note that in that paper, the emphasis is on the mathematical
aspects of dealing with order constraints, not on solving
pure strip-packing instances efficiently, as is the case in
this paper.)

For the benefit of the reader, a concise description of this approach
is contained in the following Section~\ref{sec:opp}.

\bigskip
\section{Solving Unconstrained Orthogonal Packing Problems}
\label{sec:opp}
\subsection{A General Framework}

If we have an efficient method for solving OPPs, we can also solve
SPPs by using a binary search.  However, deciding the
existence of a feasible   packing
is a hard problem in higher dimensions, and proposed methods suggested
by other authors \cite{Bea85,HC95} have been of limited success.

\begin{figure}[htbp]
  \begin{center}
    \includegraphics[width=\columnwidth]{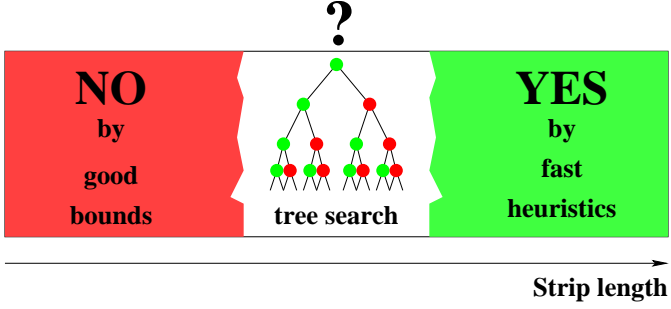}
    \caption{The basic idea of our binary search.}
    \label{fig:binary}
  \end{center}
\end{figure}

Our framework uses a combination of different approaches to overcome
these problems, see Figure~\ref{fig:binary}:
\begin{enumerate}
\item Try to disprove the existence of a packing by 
    classes of lower bounds on the necessary size.
\item In case of failure, try to find a feasible packing by using fast
  heuristics.
\item If the existence of a packing is still unsettled, start an
  enumeration scheme in form of a branch-and-bound tree search.
\end{enumerate}

By developing good new bounds for the first stage, we have been able
to achieve a considerable reduction of the number of cases where a
tree search needs to be performed.  (Mathematical details for this
step are described in \cite{IPCO,pack2}.)  However, it is clear that
the efficiency of the third stage is crucial for the overall running
time when considering difficult problems. Using a purely geometric
enumeration scheme for this step by trying to build a partial
arrangement of boxes is easily seen to be immensely time-consuming. In
the following, we describe a purely combinatorial characterization of
feasible packings that allows to perform this step more efficiently.

\subsection{Packing Classes}
\label{sec:packing-classes}

\begin{figure}[htbp]
  \begin{center}
    \includegraphics[width=.9\columnwidth]{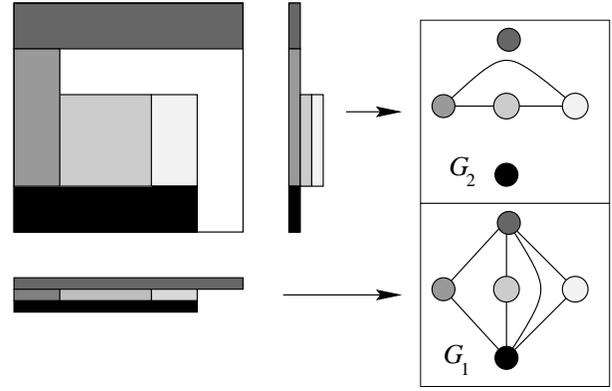}
    \caption{The projections of the boxes onto the
      coordinate axes define interval graphs (here in 2D:
      \boldmath{$G_1$} and \boldmath{$G_2$}).}
    \label{fig:intervgra}
  \end{center}
\end{figure}

Consider a feasible packing in $d$-dimensional space, and project the
boxes onto the $d$ 
  coordinate axes.  This converts the one
$d$-dimensional arrangement into $d$ one-dimensional ones (see
Figure~\ref{fig:intervgra} for an example in $d=2$).  By disregarding
the exact coordinates of the resulting intervals in direction $i$ and
only considering their intersection properties,
  we get the \emph{component graph}
$G_i=(V,E_i)$: Two boxes $u$ and $v$ are connected by an edge in
$G_i$, iff their projected intervals in direction $x_i$ have a
non-empty intersection.
   By definition, these graphs are
\emph{interval graphs}. This class of graphs has been
studied intensively in graph theory (see \cite{GOL80,RHM}), and it
has a number of very useful algorithmic properties.

Considering sets of $d$ component graphs $G_i$ instead of complicated
geometric arrangements has some clear advantages (algorithmic
implications for our specific purposes are discussed further down).
It is not hard to check that the following three conditions must be
satisfied by all $d$-tuples of graphs $G_i$ that are constructed from
a feasible packing:

\begin{enumerate}[C1:]
\item $G_i$ is an interval graph, $\forall i \in \{1,\cdots,d\}$.
\item Any independent set $S$ of $G_i$ is $i$-admissible, $\forall i
  \in \{1,\cdots,d\}$, i.e., $w_i(S) = \sum_{v \in S} w_i(v) \leq
  h_i$, because all boxes in $S$ must fit into the container in the
  $i$th dimension.
\item $\cap_{i=1}^{d} E_i = \emptyset$. In other words, there must be
  at least one dimension in which the corresponding boxes do not
  overlap.
\end{enumerate}

A $d$-tuple of component graphs satisfying these necessary conditions
is called a \emph{packing class}.  The remarkable property (proven in
\cite{Sch97,pack1})
  is that these three conditions are also sufficient for the
existence of a feasible packing.

\medskip

\begin{theorem}[Fekete, Schepers]
  A set of boxes allows a feasible packing, iff there
  is a a packing class, i.\,e., 
 a $d$-tuple of graphs $G_i=(V,E_i)$ that satisfies the
  conditions {\bf C1, C2, C3}.
\end{theorem}

\begin{figure}[htbp]
  \begin{center}
    \includegraphics[width=\columnwidth]{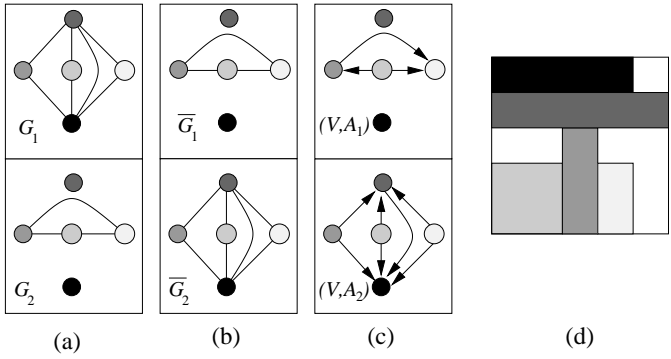}
    \caption{(a) A two-dimensional packing class.
      (b) The corresponding comparability graphs. (c) The transitive
      orientations. (d) A feasible packing
      corresponding to the orientation.}
    \label{fig:orient}
  \end{center}
\end{figure}

\medskip
This allows it to consider only packing
classes in order to decide the existence of a feasible packing, and to
disregard most of the geometric information.
See Figure~\ref{fig:orient} to see how a packing class gives rise
to a feasible packing; note that this packing is not identical to
the one in Figure~\ref{fig:intervgra}. (In fact, there are
many possible packings for a packing class, see the following subsection
and Figure~\ref{fig:orient}.)

\subsection{Solving OPPs}
\label{sec:solving-opps}

Our search procedure works on packing classes, i.e., $d$-tuples of
component graphs with the properties C1, C2, C3.  Because each packing
class represents not only a single packing but a whole family of
equivalent packings, we are effectively dealing with more than one
possible candidate for an optimal packing at a time.  (The reader may
check for the example in Figure~\ref{fig:intervgra} that there are 36
different feasible packings that correspond to the same packing
class.)

\begin{figure}[htbp]
  \begin{center}
    \includegraphics[width=.565\columnwidth]{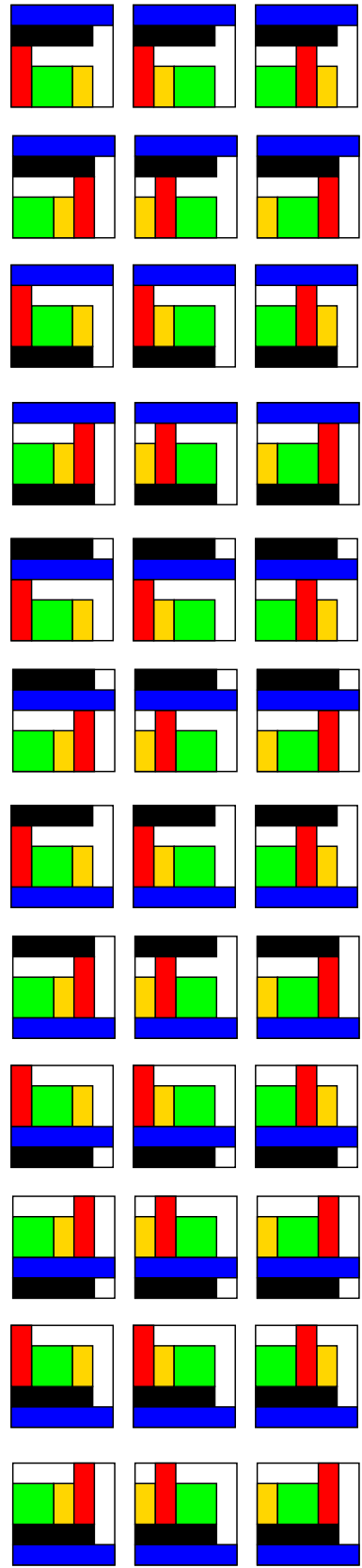}
    \caption{All shown 36 packings correspond to the component graphs 
$G_1$ and $G_2$ that are shown in Figure~\ref{fig:intervgra}.}
    \label{fig:patterns}
  \end{center}
\end{figure}

For finding an optimal packing, we use a branch-and-bound
approach. The search tree is traversed by
depth first search, see
\cite{pack3,Sch97} for details.  Branching is done by fixing an edge
$\{b,c\} \in E_{i}$ or $\{b,c\} \notin E_{i}$.  After each branching
step, it is checked whether one of the three conditions C1, C2, C3 is
violated; furthermore it is checked,
whether a violation can only be avoided by fixing further edges.
Testing for two of the conditions C1--C3 is easy:
  enforcing C3 is obvious; property C2 is hereditary, so
adding edges to $E_i$ later will keep it satisfied.  (Note that
computing maximum weighted cliques on comparability graphs can be done
efficiently, see \cite{GOL80}.)  In order to ensure that property C1
is not violated, we use some
graph-theoretic characterizations of interval graphs and comparability
graphs.  These characterizations are based on two forbidden
substructures (again, see \cite{GOL80} for details; the first
condition is based on the classical characterizations by
\cite{Gh62,GiHo64}: a graph is an interval graph \emph{iff} its
complement has a transitive orientation, and it does not contain any
induced chordless cycle of length 4.)
  In particular,
the following configurations have to be avoided:

\begin{enumerate}[G1:]
\item induced chordless cycles of length 4 in $E_{i}$;
\item so-called 2-chordless odd cycles in the set $\overline{E_i}$ of
  edges excluded from $E_i$ (see \cite{pack3,GOL80} for details);
\item infeasible stable sets in $E_i$.
\end{enumerate}
  Each time we detect such a fixed subgraph,
we can abandon the search on this node.  Furthermore, if we detect a
fixed subgraph, except for one unfixed edge, we can fix this edge,
such that the forbidden subgraph is avoided.

Our experience shows that in the considered examples
these conditions are already useful when only small subsets of edges
have been fixed, because by excluding small sub-configurations, like
induced chordless cycles of length 4, each branching step triggers a
cascade of more fixed edges.


\section{Computational Results}

\begin{figure}
\begin{center}
\begin{codebox}
\Procname{$\proc{DefragmentModuleLayout}()$}
\li $LB \gets \proc{CalculateLowerBound()}$
\li $UB \gets \proc{CalculateUpperBound()}$
\li \While $LB \neq UB$ \Do
\li     $W \gets LB + \left\lfloor\frac{LB + UB}{2}\right\rfloor$
\li     \If $\proc{SolveOPP(W)}$ \Then
\li         $LB \gets W$
\li     \Else
\li         $UB \gets W$
\end{codebox}
\end{center}
\caption{The binary search algorithm for determining an optimal module layout. In this algorithm the OPP as described in section~\ref{sec:opp} is solved repeatedly to determine if all modules fit in a strip of width $W$. This search is iterated until an optimal solution is found.}
\label{fig:spc}
\end{figure} 

We have used our implementation for the OPP (as described in the
previous section) as a building block for our new strip-packing code.
To allow for a later implementation of the strip-packing code on the
MicroBlaze cores we have used very simple lower and upper bounds to
restrict the binary search interval: Let $I$ denote the indices of the
modules present on the FPGA. Then the lower bound for the number of
columns $W_L$ we used is given by
\[W_L=\left\lceil\frac{\sum_{i \in I}w_ih_i}{H}\right\rceil.\]   
The upper bound is computed as the minimum of the three shelf-packing
heuristics next-fit-decreasing, first-fit-decreasing and
best-fit-decreasing \cite{bs-satdp-83}. These heuristics partition the
strip into shelves. A new shelf of height $h_j$ is created if there is
no shelf in which the module $j$ can be placed. If the module can be
placed in more than one shelf the shelf is picked according to the
next-fit, first-fit, or best-fit strategy respectively.

Based on these bounds the algorithm performs a binary search until an
optimal soution is found. The algorithm is outlined in
Figure~\ref{fig:spc}.

\begin{figure}[htb]
  \centering 
  \includegraphics[width=\columnwidth]{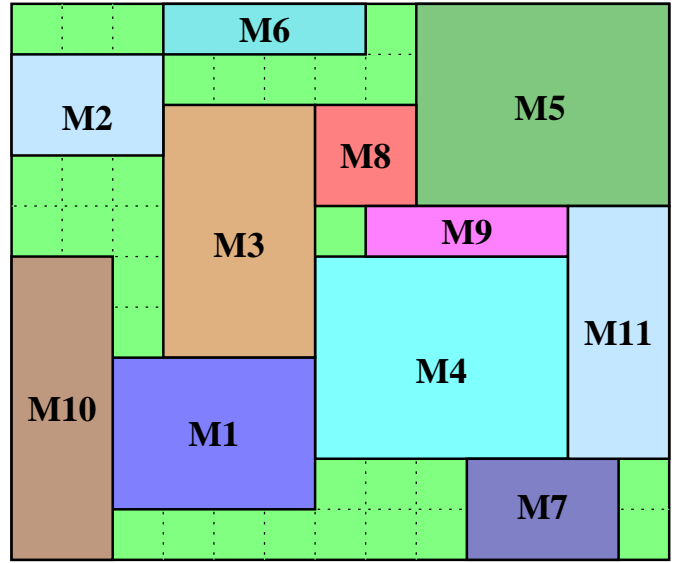}
  \caption{The FPGA before defragmentation. Even though the remaining free space is 30 reconfigurable units (RFUs), the maximal free rectangle of dimension $7\times 1$ has only 7 RFUs. Note that there is no free column.}
  \label{fig:defrag1}
\end{figure}

We have benchmarked our code against a set of 10 instances.
Considering our multimedia scenario, we have constrained different IP
cores like MPEG2 decoders, MP3 decoders, MicroBlaze core, interface
modules like CAN, CardBus, etc. to rectangular shape. We consider one
busy time, where many modules are placed and removed from the FPGA.
The placement strategy we used was LIF. For the removal of the FPGAs
we used the least-recently-used (LRU) strategy. The result is shown in
Figure~\ref{fig:defrag1}. This is followed by the removal of some
randomly selected modules. For these instances we report the maximal
free rectangle and the number of free columns before and after
defragmentation. On an Intel Pentium IV clocked at 3GHz the running
time was less than 0.5 s for each scenario.

As shown in Table~\ref{tab:results} defragmentation increases the area
of the maximal free rectangle and the number of free columns in all of
the 10 scenarios. The smallest increase in area can be seen in
scenarios E and J. Here a factor of 1.4 is obtained. In scenarios A
and C an increase of area of the maximal rectangle reaches its maximum
with a factor of 3.1. On average, the area of the maximal free
rectangle is increased by a factor of 2.2. The number of free columns
grows at least two and by at most six. The average increase of free
columns is 4.2.

  
\begin{figure}[htb]
  \centering 
  \includegraphics[width=\columnwidth]{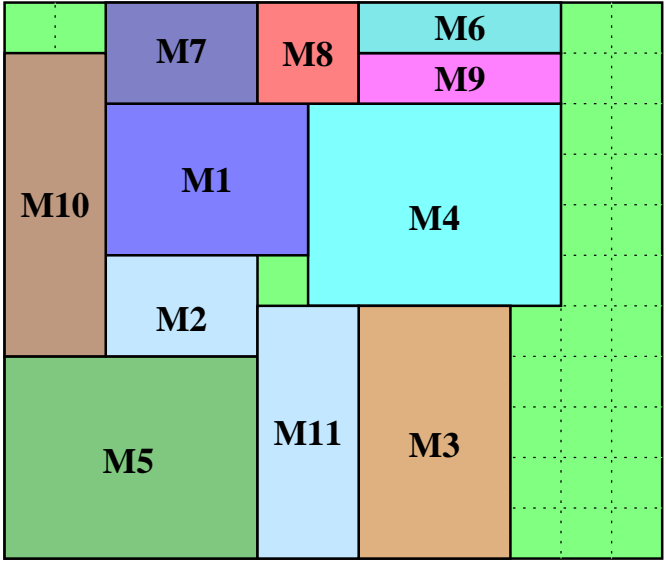}
  \caption{The same FPGA as in Figure~\ref{fig:defrag1} after defragmentation. The remaining free space is 30 reconfigurable units (RFUs). Now the maximal free rectangle is of dimension $2\times 11$ has 22 RFUs. The number of free columns is 2.}
  \label{fig:defrag2}
\end{figure}

\begin{figure}[htb]
  \centering 
  \includegraphics[width=\columnwidth]{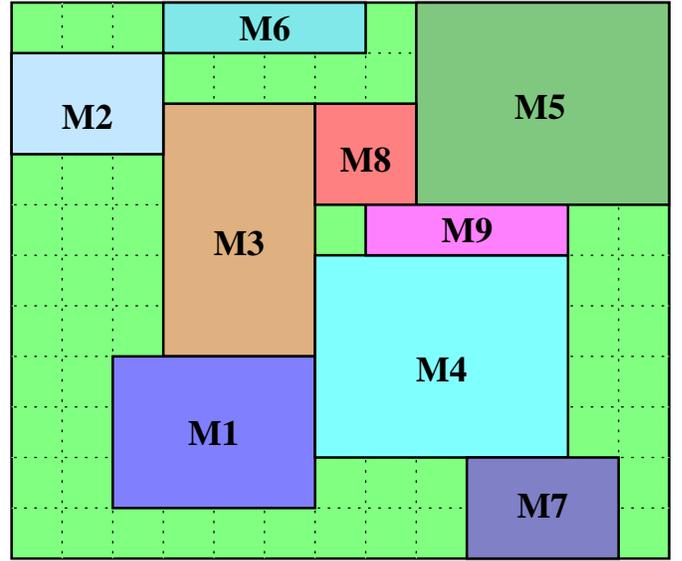}
  \caption{The same FPGA as in Figure~\ref{fig:defrag1}. Modules M10 and M11 have been removed due to a low usage count. The remaining free space is now 52 RFUs. The largest free rectangle has dimension $2\times 8$ and 16 RFUs. There still is no free column.}
  \label{fig:defrag3}
\end{figure}

\begin{figure}[htb]
  \centering 
  \includegraphics[width=\columnwidth]{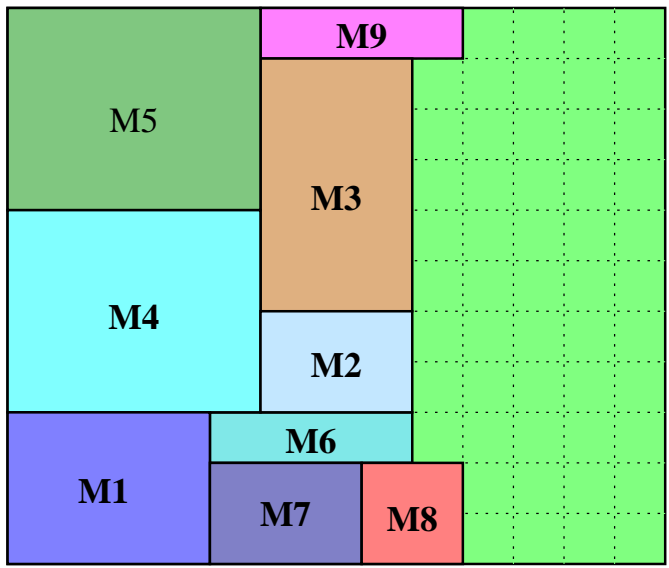}
  \caption{The same FPGA as in Figure~\ref{fig:defrag2} after succesful defragmentation. The free space of 52 RFUs is the same as before. The largest free rectangle has grown to dimension $4 \times 11$ and contains 44 RFUs. Now there are 4 free columns.}
  \label{fig:defrag4}
\end{figure}


\begin{table*}
\begin{center}
\begin{tabular}{|c|c|c|c|c|c|c|}
\hline
         &       &            & \multicolumn{2}{l|}{Before defragmentation} & \multicolumn{2}{l|}{After defragmentation} \\
Scenario & $|I|$ & Free space & Max. rectangle & Free columns & Max. rectangle & Free columns \\
\hline
A & 11 & 30 & $7 \times  1$ & 0 & $2 \times 11$ & 2 \\ 
B &  9 & 52 & $2 \times  8$ & 0 & $4 \times 11$ & 4 \\
C &  9 & 70 & $3 \times  7$ & 0 & $6 \times 11$ & 6 \\
D &  9 & 42 & $4 \times  4$ & 0 & $3 \times 11$ & 3 \\
E &  6 & 83 & $6 \times  8$ & 0 & $6 \times 11$ & 6 \\
F &  6 & 54 & $8 \times  2$ & 0 & $4 \times 11$ & 4 \\
G &  5 & 76 & $6 \times  4$ & 2 & $6 \times 11$ & 6 \\
H &  6 & 53 & $3 \times 11$ & 3 & $4 \times 11$ & 7 \\
I &  5 & 87 & $9 \times  6$ & 1 & $7 \times 11$ & 7 \\
J &  6 & 42 & $3 \times  8$ & 0 & $3 \times 11$ & 3 \\
\hline
\end{tabular}
\end{center}
\caption{Results for ten different scenarios, based on Figure~\ref{fig:defrag1}. 
The solution for scenario A is shown in Figure~\ref{fig:defrag2}, the others 
have modules removed from the FPGA. Scenario B is shown in 
Figure~\ref{fig:defrag3}, its solution in Figure~\ref{fig:defrag4}. 
The next columns show the number of placed modules, the total free space, the maximal free rectangle, and the number of free columns before defragmentation. The final columns show results after defragmentation.}\label{tab:results}
\end{table*}



\section{Conclusions}

We have shown that mixing online and offline strategies
can improve the overall reconfiguration process in partial
reconfiguration. Especially for FPGAs with partial reconfiguration
restricted to columnwise reconfiguration, a defragmentation strategy
as proposed in this paper helps to reduce the interference with other
modules.

There are many possible extensions to our approach. We list two of them
explicitly:
\begin{enumerate}
\item {\em Malleable modules:} Tools for automatic synthesis normally do
  not create modules with rectangular shape. 
  Instead, width
  and height of the modules can be chosen freely within certain technical 
  bounds. This gives more room
  for the optimization in the defragmentation process. In a
  mathematical context this model would be called a {\em class strip
  packing problem}: Given a set of modules that has to be placed on a
  chip as to minimize the total number of columns used, choose for
  each module from a certain set of module realizations and try to
  find a placement. 
  
  If the width and height of the modules can be chosen freely this
  problem is known as {\em strip packing with modifiable boxes}. In an
  offline setting this problem can be trivially solved by applying
  once the volume lower bound as described above and then setting the
  height of each box to this value. In \cite{i-ospmb-01} the author
  gives a 4-competitive online algorithm for the problem and shows
  that no online algorithm can do better than 1.73.
\item {\em Fixed modules:} In most FPGA designs, pins of the FPGA are
  hard-wired. In this setting it may be unavoidable to fix a placement
  of the respective interface modules in close proximity to their IO
  pins. When this is the case, the defragmentation problem is no
  longer a strip-packing problem.  Freeing as many columns as possible
  can be achieved by placing other modules above or below the
  interface modules and not just as far as possible to the left.
\end{enumerate}
We are optimistic that our general approach will allow some progress
on these problem classes.

\section*{Acknowledgments}
We are extremely grateful to J{\"o}rg Schepers for letting us continue
the work with the packing code that he started as part of his thesis,
and for several helpful hints, despite of his departure to industry.

This research has been supported by the Deutsche
Forschungsgemeinschaft (DFG) as part of the project ``ReCoNodes'',
grant numbers FE407/8-1 and \mbox{TE163/11-1}.


\bibliographystyle{IEEEtranS}
\bibliography{Defragmentation}

\begin{thebibliography}{10}
\providecommand{\url}[1]{#1}
\csname url@rmstyle\endcsname
\providecommand{\newblock}{\relax}
\providecommand{\bibinfo}[2]{#2}
\providecommand\BIBentrySTDinterwordspacing{\spaceskip=0pt\relax}
\providecommand\BIBentryALTinterwordstretchfactor{4}
\providecommand\BIBentryALTinterwordspacing{\spaceskip=\fontdimen2\font plus
\BIBentryALTinterwordstretchfactor\fontdimen3\font minus
  \fontdimen4\font\relax}
\providecommand\BIBforeignlanguage[2]{{%
\expandafter\ifx\csname l@#1\endcsname\relax
\typeout{** WARNING: IEEEtran.bst: No hyphenation pattern has been}%
\typeout{** loaded for the language `#1'. Using the pattern for}%
\typeout{** the default language instead.}%
\else
\language=\csname l@#1\endcsname
\fi
#2}}

\bibitem{swp-osrep-04}
C.~Steiger, H.~Walder, and M.~Platzner, ``Operating systems for reconfigurable
  embedded platforms: Online scheduling of real-time tasks,'' \emph{IEEE
  Transaction on Computers}, vol.~53, no.~11, pp. 1393--1407, 2004.

\bibitem{demss_dstpr-00}
O.~Diessel, H.~{ElGindy}, M.~Middendorf, H.~Schmeck, and B.~Schmidt, ``Dynamic
  scheduling of tasks on partially reconfigurable {FPGAs},'' \emph{IEE
  Proceedings -- Computers and Digital Techniques}, vol. 147, no.~3, pp.
  181--188, May 2000.

\bibitem{x-vpfcd-04}
\emph{Virtex-{II} platform {FPGAs}: Complete data sheet}, XILINX Inc., June
  2004.

\bibitem{bmkat-dnacr-04}
C.~Bobda, M.~Majer, D.~Koch, A.~Ahmadinia, and J.~Teich, ``A dynamic {NoC}
  approach for communication in reconfigurable devices,'' in \emph{Proceedings
  of International Conference on Field-Programmable Logic and Applications
  (FPL)}, ser. Lecture Notes in Computer Science (LNCS), vol. 3203.\hskip 1em
  plus 0.5em minus 0.4em\relax Antwerp, Belgium: Springer, Aug. 2004, pp.
  1032--1036.

\bibitem{sched}
E.~L. Lawler, J.~K. Lenstra, A.~H.~G. {Rinooy Kan}, and D.~B. Shmoys,
  ``Sequencing and scheduling: Algorithms and complexity,'' in \emph{Logistics
  of Production and Inventory}, ser. Handbooks in Operations Research and
  Management, vol.~4, S.~C. Graves, A.~H.~G. {Rinnooy Kan}, and P.~H. Zipkin,
  Eds.\hskip 1em plus 0.5em minus 0.4em\relax North--Holland, Amsterdam, 1993,
  pp. 445--522.

\bibitem{at-sopxf-03}
A.~Ahmadinia and J.~Teich, ``Speeding up online placement for {XILINX} {FPGA}s
  by reducing configuration overhead,'' in \emph{Proceedings of the {IFIP}
  International Conference on VLSI-SOC}.\hskip 1em plus 0.5em minus 0.4em\relax
  Darmstadt, Germany: IFIP, Dec. 2003, pp. 118--122.

\bibitem{GJ79}
M.~R. Garey and D.~S. Johnson, \emph{Computers and Intractability: A Guide to
  the Theory of NP-Completeness}.\hskip 1em plus 0.5em minus 0.4em\relax New
  York: Freeman, 1979.

\bibitem{GRO80}
M.~Gr{\"o}tschel, ``On the symmetric travelling salesman problem: solution of a
  120-city problem,'' \emph{Mathematical Programming Study}, vol.~12, pp.
  61--77, 1980.

\bibitem{13509}
D.~Applegate, R.~Bixby, V.~Chv{\'a}tal, and W.~Cook, ``On the solution of
  traveling salesman problems,'' \emph{Documenta Mathematica Journal der
  Deutschen Mathematiker-Vereinigung}, vol. ICM III, pp. 645--656, 1998.

\bibitem{ESA}
S.~P. Fekete and J.~Schepers, ``A new exact algorithm for general orthogonal
  d-dimensional knapsack problems,'' in \emph{Algorithms -- ESA '97}, vol.
  1284, Springer Lecture Notes in Computer Science, 1997, pp. 144--156.

\bibitem{pack1}
------, ``A combinatorial characterization of higher-dimensional orthogonal
  packing,'' \emph{Mathematics of Operations Research}, vol.~29, pp. 353--368,
  2004.

\bibitem{Bea85}
J.~E. Beasley, ``An exact two-dimensional non-guillotine cutting tree search
  procedure,'' \emph{Operations Research}, vol.~33, pp. 49--64, 1985.

\bibitem{HC95}
E.~Hadjiconstantinou and N.~Christofides, ``An exact algorithm for general,
  orthogonal, two-dimensional knapsack problems,'' \emph{European Journal of
  Operations Research}, vol.~83, pp. 39--56, 1995.

\bibitem{mmv-easpp-03}
S.~Martello, M.~Monaci, and D.~Vigo, ``An exact approach to the strip-packing
  problem,'' \emph{{INFORMS} Journal on Computing}, vol.~15, no.~3, pp.
  310--319, 2003.

\bibitem{pack2}
S.~P. Fekete and J.~Schepers, ``A general framework for bounds for
  higher-dimensional orthogonal packing problems,'' \emph{Mathematical Methods
  of Operations Research}, vol.~60, pp. 311--329, 2004.

\bibitem{pack3}
------, ``An exact algorithm for higher-dimensional packing,'' \emph{Operations
  Research}, to appear.

\bibitem{tfs-ohrt-01}
J.~Teich, S.~P. Fekete, and J.~Schepers, ``Optimal hardware reconfiguration
  techniques,'' \emph{Journal of Supercomputing}, vol.~19, pp. 57--75, 2001.

\bibitem{IPCO}
S.~P. Fekete and J.~Schepers, ``New classes of lower bounds for bin packing
  problems,'' in \emph{Integer Programming and Combinatorial Optimization
  (IPCO'98)}, ser. Springer Lecture Notes in Computer Science, vol. 1412, 1998,
  pp. 257--270.

\bibitem{fkt-mdpoc-01}
S.~P. Fekete, E.~K{\"o}hler, and J.~Teich, ``Multi-dimensional packing with
  order constraints,'' in \emph{Proceedings 7th International Workshop on
  Algorithms and Data Structures}, ser. Lecture Notes in Computer Science, vol.
  2125.\hskip 1em plus 0.5em minus 0.4em\relax Springer-Verlag, 2001, pp.
  300--312.

\bibitem{GOL80}
M.~C. Golumbic, \emph{Algorithmic graph theory and perfect graphs}.\hskip 1em
  plus 0.5em minus 0.4em\relax New York: Academic Press, 1980.

\bibitem{RHM}
R.~H. M{\"o}hring, ``Algorithmic aspects of comparability graphs and interval
  graphs,'' in \emph{Graphs and Order}, I.~Rival, Ed.\hskip 1em plus 0.5em
  minus 0.4em\relax D. Reidel Publishing Company, Dordrecht, 1985, pp. 41--101.

\newpage
\bibitem{Sch97}
J.~Schepers, ``Exakte {A}lgorithmen f{\"u}r orthogonale {P}ackungsprobleme,''
  Uni\-ver\-si\-t{\"a}t K{\"o}ln, Tech. Rep. 97-302, Doctoral thesis, 1997.

\bibitem{Gh62}
A.~Ghouil\`{a}-Houri, ``Caract\'{e}rization des graphes non orient\'{e}s dont
  on peut orienter les arr\^{e}tes de mani\`{e}re \`{a} obtenir le graphe d'une
  relation d'ordre,'' \emph{C.R. Acad. Sci. Paris}, vol. 254, pp. 1370--1371,
  1962.

\bibitem{GiHo64}
P.~C. Gilmore and A.~J. Hoffmann, ``A characterization of comparability graphs
  and of interval graphs,'' \emph{Canadian Journal of Mathematics}, vol.~16,
  pp. 539--548, 1964.

\bibitem{bs-satdp-83}
B.~S. Baker and J.~S. Schwarz, ``Shelf algorithms for two-dimensional packing
  problems,'' \emph{SIAM Journal on Computing}, vol.~12, no.~3, pp. 508--525,
  1983.

\bibitem{i-ospmb-01}
C.~Imreh, ``Online strip packing with modifiable boxes,'' \emph{Operations
  Research Letters}, vol.~29, no.~2, pp. 79--85, 2001.

\end{thebibliography}

\end{document}